\documentclass[reqno]{article}

\usepackage{graphicx}
\usepackage{amsmath,amsthm,amssymb}

\chardef\bslash=`\\ % p. 424, TeXbook
\theoremstyle{definition}

\theoremstyle{remark}

\newcommand{\eval}[2][\right]{\relax
  \ifx#1\right\relax \left.\fi#2#1\rvert}

\begin{document}
\title{\bf{Dispersion and collapse of wave maps}}

\author{Piotr Bizo\'n\footnotemark[1]{},
  Tadeusz Chmaj\footnotemark[2]{},  and Zbislaw
  Tabor\footnotemark[1]{}\\{}\\
  \footnotemark[1]{} \small{\textit{Institute of Physics,
   Jagellonian University, Krak\'ow, Poland}}\\
   \footnotemark[2]{} \small{\textit{Institute of Nuclear Physics, Krak\'ow,
   Poland}}}
\date{December 12, 1999}
\maketitle
\begin{abstract}
\noindent We study numerically the Cauchy problem for equivariant
wave maps from  $3+1$ Minkowski spacetime into the 3-sphere. On
the basis of
 numerical evidence combined with stability analysis of self-similar solutions
we formulate two conjectures. The first conjecture states that
singularities which are produced in the evolution of sufficiently
large initial data are approached in a universal manner given by the
profile of a stable self-similar solution. The second conjecture
states that the codimension- one stable manifold of a self-similar
solution with exactly one instability determines the threshold of
singularity formation for a large class of initial data. Our results
can be considered as a toy-model for some aspects of the critical
behavior in formation of black holes.
\end{abstract}
\section{Introduction}
Let $M$ be a spacetime with metric $\eta$ and $N$ be a Riemannian
manifold with metric $g$.  The wave  map $U: M \rightarrow N$ is
defined as a critical point of the action
\begin{equation}
S(U) = \int_{M} g_{AB}\, \partial_a U^A \partial_b U^B  \eta^{ab} \,
dV_M \, .
\end{equation}
The associated Euler-Lagrange equations constitute the system of
semilinear wave equations
\begin{equation}\label{wmap}
\partial^a\partial_a U^A + \Gamma_{BC}^A(U) \partial_aU^B \partial^a
U^C=0,
\end{equation}
where $\Gamma$'s are the Christoffel symbols of the target metric
$g$.

 The recent surge of interest in wave maps (known in the physics literature
 as $\sigma$-models) stems from the fact that they provide an
attractive toy-model for more complicated relativistic field
equations. In particular they share some features with the Einstein
equations so  understanding the problems of global existence and
formation of singularities for wave maps may shed some light on the
analogous, but much more difficult, problems in general relativity.
Having this analogy in mind we have studied the  Cauchy problem for
Eq.(2) in the case where the domain space is  $3+1$ Minkowski
spacetime, $M=\mathbb{R}^{3+1}$, and the target space is the
3-sphere, $N=S^3$. Although our primary motivation was an attempt to
get insight into some aspects of critical behavior in gravitational
collapse, we think that our results are interesting in their own right. In
this paper we restrict attention to equivariant maps. In polar
coordinates on $\mathbb{R}^{3+1}$ and $S^3$ the respective metrics
are
\begin{equation}
\eta= -dt^2 + dr^2 + r^2 d\omega^2, \quad g = du^2 + \sin^2(u)
d\Omega^2,
\end{equation}
 where $d\omega^2$ and $d\Omega^2$
are the standard metrics on $S^2$. Equivariant maps have the form
\begin{equation}
U(t,r,\omega)= (u(t,r),\Omega(\omega)), \end{equation} where $\Omega$
is a homogeneous harmonic polynomial of degree $l>0$. In what follows
we consider the case $l=1$, where $\Omega=\omega$ (such  maps are
called corotational). For a corotational map the Cauchy problem for
Eq.(\ref{wmap})
 reduces to
the semilinear wave equation
\begin{equation}\label{eq}
u_{tt}= u_{rr}+\frac{2}{r} u_r - \frac{\sin(2u)}{r^2}
\end{equation}
with initial data
\begin{equation}\label{data}
u(0,r)=\phi(r), \quad u_t(0,r)=\psi(r).
\end{equation}
The conserved energy associated with solutions of this equation
\begin{equation}\label{energy}
E[u]=\frac{1}{2} \int_0^{\infty} (r^2 u_t^2+ r^2 u_r^2+
  2\sin^2\!{u})\: dr
\end{equation}
is manifestly nonnegative  and scales as $E[u(x/\lambda)]=\lambda
E[u(x)]$ which means that Eq.(\ref{eq}) is supercritical (like
Einstein's equations). It is widely believed that for supercritical
equations the solutions with sufficiently small initial data exist
for all times while large data solutions develop singularities in
finite time~\cite{klainer}. In the case of (\ref{eq}) the global
existence for small (in the Sobolev space $H^k$ with sufficiently
large $k$) data was proved by Kovalyov~\cite{kov} and
Sideris~\cite{sid}. For large data there are no rigorous results,
however it is known that there exist smooth data which lead to blowup
in finite time. An example of such data  is due to Shatah who showed
that (\ref{eq}) admits a self-similar solution
$u(t,r)=f_0(\tfrac{r}{T-t})$ which is perfectly smooth for $t<T$ but
breaks down at $t=T$. Turok and Spergel~\cite{turok} found this
solution in closed form $f_0=2 \arctan(\frac{r}{T-t})$ so in the
following we will refer to $f_0$ as the TS solution.

On the basis of our numerical simulations we conjecture that the
example of blowup given by Shatah is generic. By this we mean that
there is a large open set of initial data which blow up in a finite
time $T$ and the asymptotic shape of solutions near the blowup point
($r=0$) approaches $f_0(\frac{r}{T-t})$ as $t\rightarrow T^-$. In
this sense the blowup can be considered as local convergence to the
TS solution $f_0$. Actually, our failed efforts to produce a
non-self-similar singularity lead us to suspect that the blowup is
universally self-similar.  Note that the self-similarity of blowup
excludes a concentration of energy at the singularity and suggests
that the solutions can be continued beyond the blowup time in an
almost continuous fashion. We must admit that this aspect of
singularity formation for wave maps is somewhat disappointing from
the standpoint of modeling the formation of energy trapping
singularities (like black holes).

Whenever the singularities develop from some but not all data, there
arises a natural question of determining the threshold of singularity
formation. We investigated this issue using a basic technique of
evolving various one-parameter families of initial data which
interpolate between global existence (dispersion) and blowup. Along
each such family there exists a point (critical initial data) which
separates the two regimes. We show that the critical initial data
blow up in a finite time $T$ and the asymptotic shape of solutions
near the blowup point ($r=0$) approaches $f_1(\frac{r}{T-t})$, a
self-similar solution with one unstable mode. The marginally critical
data approach $f_1$ for intermediate times but eventually the
unstable mode becomes dominant and ejects the solutions towards
dispersion or stable blowup (that is, $f_0$). Thus, we conjecture
that the codimension-one stable manifold of the solution $f_1$ plays
the role of the threshold of singularity formation for a large set of
initial data.

The threshold behavior in our model is similar to the type II
critical behavior in gravitational collapse (see~\cite{gund} for the
recent review) where self-similar solutions of Einstein's equations
(continuous or discrete, depending on a model) sit at the threshold
of formation of a black hole. There are also many parallels  between
our results and the work of Brenner et al~\cite{brenner} on the
chemotaxis equation. All that suggests that self-similar behavior at
the threshold of singularity formation is a common feature for
evolutionary partial differential equation.

The rest of the paper in organized as follows. In the next two
sections we discuss some special solutions of (\ref{eq}) which are
potential candidates for attractors. In Section~2 we analyze
self-similar solutions and their linear stability.  Section~3 is
devoted to static solutions. In Section~4 we describe the results of numerical simulations
and document the numerical evidence behind the two conjectures formulated above.

\section{Self-similar solutions}
Note that Eq.(\ref{eq}) is invariant under dilations: if $u(t,r)$ is
a solution, so is $u_a(t,r)=u(a t,a r)$. It is thus natural
  to look for self-similar
solutions of the form
\begin{equation}\label{ansatz}
u(t,r)=f\left(\frac{r}{T-t}\right), \end{equation}
 where $T$ is a
positive constant.  Substituting the ansatz (\ref{ansatz}) into
(\ref{eq}) we obtain the ordinary differential equation
\begin{equation}\label{ss}
f''+\frac{2}{\rho} f' -\frac{\sin(2f)}{\rho^2 (1-\rho^2)} = 0,
\end{equation} where $\rho=r/(T-t)$ and $'=d/d \rho$. For $t<T$ we have $0
\leq \rho< \infty$.

It is sufficient to consider equation (\ref{ss}) only inside the past
light cone of the point $(T,0)$, that is for $\rho \in [0,1]$. The
regularity of solutions at the endpoints of this interval enforces
the following behavior
\begin{equation}\label{r=0}
  f(\rho) \sim a \rho \quad \mbox{as} \quad \rho \rightarrow 0,
\end{equation}
and
\begin{equation}\label{r=1}
  f(\rho) \sim \frac{\pi}{2} + b (1-\rho) \quad \mbox{as}
  \quad \rho\rightarrow 1,
\end{equation}
where $a$ and $b$ are arbitrary constants. At each endpoint the
parameters $a$ and $b$ determine unique local solutions. One can show
that there is a countable sequence of pairs $(a_n,b_n)$ for which the
corresponding solutions, denoted by $f_n(\rho)$, are globally regular
in the sense that they satisfy both boundary conditions (\ref{r=0})
and (\ref{r=1}) and are smooth for $\rho \in (0,1)$.
\begin{figure}[!ht]
\centering
\includegraphics[width=0.8\textwidth]{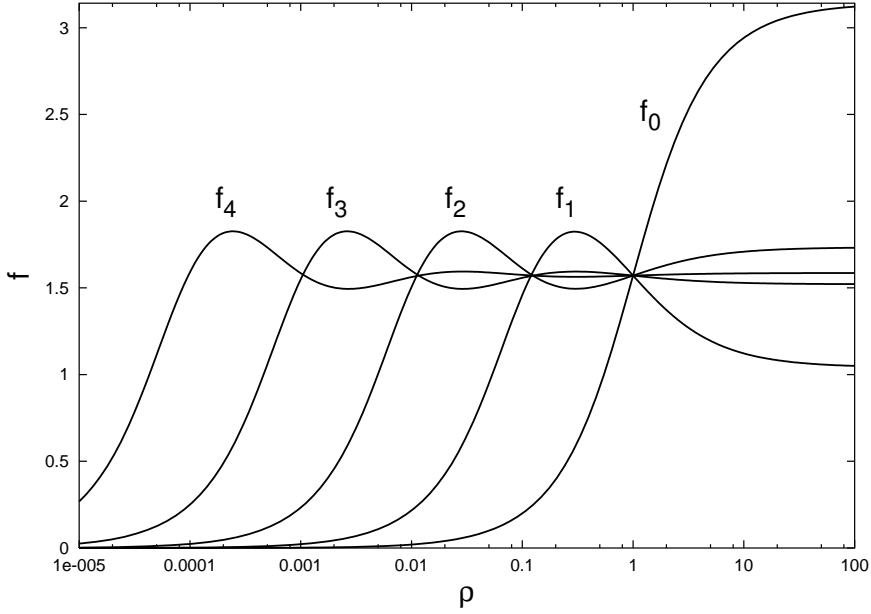}
\caption{The first five self-similar solutions.}
\end{figure}
\begin{table}[h]
  \centering
  \begin{tabular}{|c|c|c|c|c|c|} \hline
    $n$ & 0 & 1 & 2 & 3 & 4 \\ \hline
    $a_n$ & 2 & 21.757413 & 234.50147 & 2522.0683 & 27113.388 \\
    $b_n$ & 1 & -0.305664 & 0.0932163  & -0.0284312 & 0.0086717 \\ \hline
  \end{tabular}
  \caption{The parameters of  solutions shown in Fig.~1.}
\end{table}

 These solutions
can be smoothly extended for $\rho>1$ by solving (9) with the initial
condition (11). One can show that the asymptotic behavior for
$\rho\rightarrow \infty$ is
\begin{equation}\label{asym}
f_n(\rho) \sim c_n +\frac{d_n}{\rho} + O(\frac{1}{\rho^2}),
\end{equation}
where $c_n \rightarrow \pi/2$ as $n \rightarrow \infty$.
 The countable family $f_n$ was discovered numerically
by \"Aminneborg and Bergstr\"{o}m~\cite{ab}; recently  its
existence was proven rigorously via a shooting argument~\cite{ja}.
The integer index $n=0,1,\ldots$ denotes the number of intersections
of $f_n(\rho)$ with the line $f=\pi/2$ on the interval $\rho\in[0,1)$. The "ground
state" solution of this family is the TS solution $f_0=2
\arctan(\rho)$.
 The solutions $f_n$ with $n>0$ can be
obtained numerically by a standard shooting-to-a-fitting-point
technique, that is by integrating equation (9) away from the singular
points $\rho=0$ and $\rho=1$ in the opposite directions with some
trial parameters $a$ and $b$ and then adjusting these parameters so
that the solution joins smoothly at the fitting point. The profiles
of solutions generated in this way (for $n \leq 4$) are shown in
Fig.~1; the corresponding parameters $(a_n,b_n)$ are given in
Table~1.

The role of self-similar solutions $f_n$ in the evolution depends
crucially on their stability with respect to small perturbations.
 In order to analyze the
linear stability of the self-similar solutions it is convenient to
define the new time coordinate $\tau=-\ln(T-t)$ and rewrite
Eq.(\ref{eq}) in terms of $\tau$ and $\rho$
\begin{equation} \label{rho-tau}
u_{\tau\tau} + u_{\tau} + 2 \rho\: u_{\rho\tau}
-(1-\rho^2)(u_{\rho\rho} +\frac{2}{\rho} u_{\rho}) +\frac{ \sin(2
u)}{\rho^2}  = 0.
\end{equation}
The standard procedure is to seek solutions of (\ref{rho-tau})
in the form $u(\tau,\rho)=f_n(\rho)+ w(\tau,\rho)$. Neglecting the
$O(w^2)$ terms we obtain a linear evolution equation for the
perturbation $w(\tau,\rho)$
\begin{equation}\label{pert}
w_{\tau\tau} + w_{\tau} + 2 \rho\: w_{\rho\tau}
-(1-\rho^2)(w_{\rho\rho} +\frac{2}{\rho} w_{\rho}) +\frac{2
\cos(2f_n)}{\rho^2} w  = 0.
\end{equation}
Substituting $w(\tau,\rho)=e^{\lambda \tau} v_{\lambda}(\rho)/\rho$
into (\ref{pert}) we get an eigenvalue problem
\begin{equation}\label{spectrum}
-(1-\rho^2) v_{\lambda}''+2 \lambda \rho v_{\lambda}'
+\lambda(\lambda-1) v_{\lambda} + \frac{2 \cos(2f_n)}{\rho^2}
v_{\lambda}=0.
\end{equation}
Near $\rho=0$ the leading behavior of solutions of (\ref{spectrum})
is $v_{\lambda}(\rho) \sim \rho^{\alpha}$ where $\alpha(\alpha-1)=2$,
so to ensure regularity we require \begin{equation}\label{rho0}
 v_{\lambda}(\rho) \sim \rho^2 \quad \mbox{as} \quad \rho\rightarrow 0.
 \end{equation}
Near $\rho=1$ the leading behavior is $v_{\lambda}(\rho) \sim
(1-\rho)^{\beta}$ where $\beta(\beta-1+\lambda)=0$. The behavior
corresponding to $\beta=1-\lambda$ is not admissible (unless
$\lambda=1$), so regular solutions must have $\beta=0$. Then we have
(up to a normalization constant)
\begin{equation}\label{rho1}
  v_{\lambda}(\rho) \sim 1 + \frac{2+\lambda (1-\lambda)}{2 \lambda}
   (1-\rho) + O((1-\rho)^2) \quad \mbox
  {as} \quad \rho\rightarrow 1.
\end{equation}
To find the eigenvalues we need to solve Eq.(\ref{spectrum}) on the
interval $\rho\in [0,1]$ with the boundary conditions (\ref{rho0})
and (\ref{rho1}). We did this  numerically (for $n\leq 4$) by
shooting the solutions from both ends and matching the logarithmic
derivative at a midpoint. Given an eigenvalue $\lambda$, the
eigenfunction $v_{\lambda}(\rho)$ can be extended for $\rho>1$ by
solving (\ref{spectrum}) with the initial condition (\ref{rho1}). Our
numerical results strongly suggest that the solution $f_n$ has
exactly $n+1$ positive eigenvalues (unstable modes). We denote them
by $\lambda^{(n)}_k$ ($k=1,\ldots,n+1$) where
$\lambda^{(n)}_1>\lambda^{(n)}_2>\ldots>\lambda^{(n)}_{n+1}=1$. For
example, for $n=1$ we have $\lambda^{(1)}_1\approx 6.333625,
\lambda^{(1)}_2=1$; for $n=2$ we have $\lambda^{(2)}_1 \approx 59.07,
\lambda^{(2)}_2\approx 6.304, \lambda^{(2)}_3=1$.
\begin{figure}[!ht]
\centering
\includegraphics[width=\textwidth]{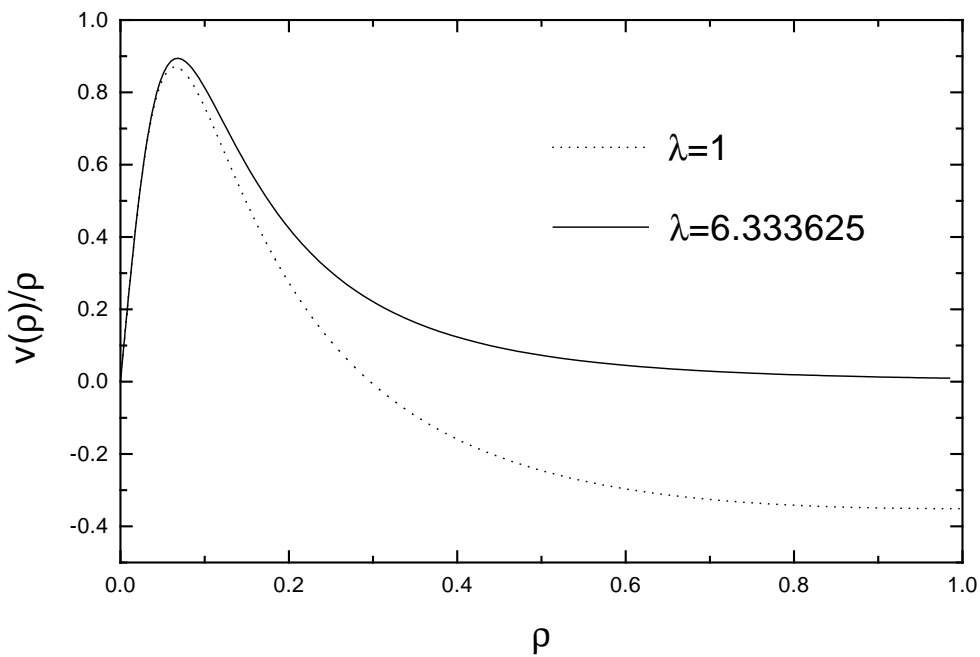}
\caption{The profiles of  unstable modes  around the solution
$f_1(\rho)$. The "real" unstable mode (solid line) corresponds to the
eigenvalue $\lambda^{(1)}_1 \approx 6.333625$. The gauge mode (dotted
line) has $\lambda^{(1)}_2=1$. For better visualization both plots
are normalized to the same slope at the origin.}
\end{figure}

  For every $n$ the
lowest positive eigenvalue $\lambda=1$ corresponds to the gauge mode
which is due to the freedom
 of choosing the blowup time $T$. To see this, consider a
solution $f_n(r/(T'-t))$. In terms of the similarity variables
$\tau=-\ln(T-t)$ and $\rho=r/(T-t)$, we have
\begin{equation}\label{gauge}
f_n\!\left(\frac{r}{T'-t}\right)=f_n\!\left(\frac{\rho}{1+\epsilon
e^{\tau}}\right) \quad \mbox{where} \quad \epsilon=T'-T.
\end{equation} In other words, each self-similar solution $f_n(\rho)$
generates the orbit of solutions of (\ref{rho-tau})
$f_n(\tfrac{\rho}{1+\epsilon e^{\tau}})$.
 It is easy to verify that the generator of this orbit
\begin{equation}\label{zeromode}
  w(\tau,\rho) = -\frac{d}{d\epsilon}\: f_n\!\left(\frac{\rho}{1+\epsilon
  e^{\tau}}\right)\Biggr\rvert_{\epsilon=0} = e^{\tau} \rho f'_n(\rho)
\end{equation}
satisfies (\ref{pert}), thus $v_n=f'_n(\rho)$ satisfies
(\ref{spectrum}) with $\lambda=1$. Note that this eigenfunction has
exactly $n$ zeros on $\rho\in (0,1)$ (since $f_n$ has $n$ extrema).
For a standard Sturm-Liouville problem this would imply the existence
of $n$ eigenvalues above $\lambda=1$. It seems feasible to prove a
similar result in the case of (\ref{spectrum}), however we do not
pursue this issue here.

\section{Static solutions}
Static solutions of Eq.(\ref{eq}) can be interpreted as spherically
symmetric harmonic maps from the Euclidean space $\mathbb{R}^3$ into
$S^3$. They satisfy the ordinary differential equation
\begin{equation}\label{static}
u''+\frac{2}{r} u' - \frac{\sin(2 u)}{r^2} =0,
\end{equation}
where now $'=d/dr$. The obvious constant solutions of (\ref{static})
are $u=0$ and $u=\pi$; geometrically these are maps into the north and the south pole of
$S^3$, respectively. The energy of these maps attains a global minimum $E=0$. Another
constant solution is the equator map $u=\pi/2$ but this solution is
singular and has infinite energy.  The scale invariance of
(\ref{static}) excludes existence of nontrivial regular solutions
with finite energy. However, there exists a regular solution with
infinite energy, denoted here by $u_S(r)$, which behaves as
\begin{equation}\label{asym}
u_S(r) \sim \begin{cases} r & \text{for $r\rightarrow 0$},\\
\frac{\pi}{2} + \frac{C}{\sqrt{r}} \sin(\frac{\sqrt{7}}{2} \ln r
+\delta) & \text{for $r\rightarrow\infty$}.
\end{cases}
\end{equation}
The existence of this solution, shown in Fig.~3, can be easily proven
using $x=\ln(r)$ which transform (\ref{static}) into a damped
pendulum equation~\cite{jk}.
\begin{figure}[!ht]
\centering
\includegraphics[width=0.8\textwidth]{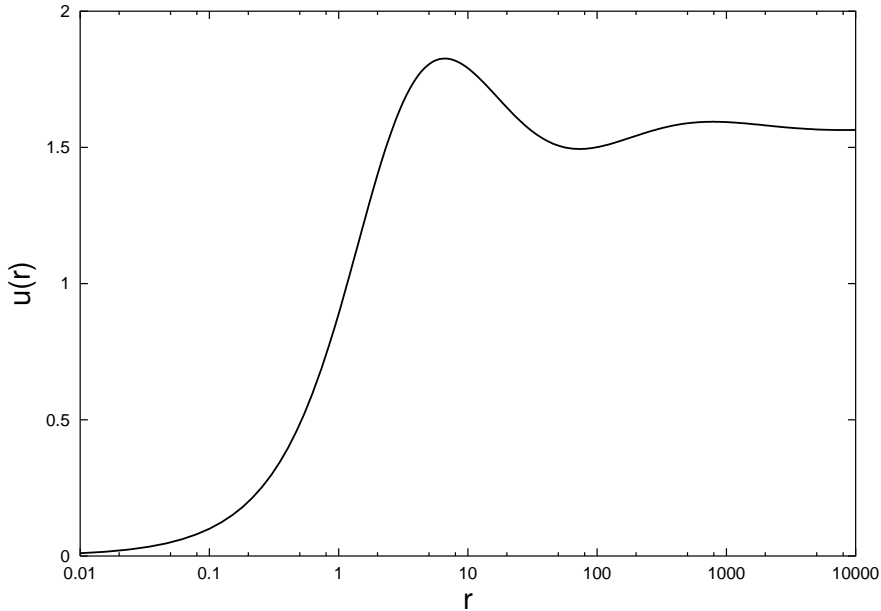}
\caption{The static solution $u_S(r)$.}
\end{figure}

Note that by dilation symmetry, the solution $u_S(r)$
 generates the orbit
of static solutions  $u_S^a(r)=u_S(a r)$.

  We consider now the  linear stability of the static solution $u_S$.
   Inserting
  $u(t,r)=u_S(r)+e^{ikt} v(r)$ into (\ref{eq}) and linearizing, we get the
  eigenvalue problem
\begin{equation}\label{pertstat}
  -v''-\frac{2}{r} v' +  V(r) v = k^2 v, \quad V(r)=\frac{2 \cos(2 u_S)}{r^2}
.
\end{equation}
If the singular $1/r^2$ part is subtracted from $V$,
then (\ref{pertstat}) becomes the p-wave radial Schr\"odinger
equation in the regular potential $V_{reg}=V(r)-2/r^2$. This
potential  has infinitely many bound states as can be shown by the
following standard argument.
 Consider the perturbation induced by the scaling
transformation
\begin{equation}\label{trick}
v(r)=\frac{d}{d a} u_S^a(r) \Bigr\rvert_{a=1} = r u_S'(r)
.
\end{equation}
This is an eigenfunction to zero eigenvalue (so called zero mode).
Since $u_S(r)$ has infinitely many extrema, the zero mode has
infinitely many nodes which implies by the standard result from
Sturm-Liouville theory  that there are infinitely many negative
eigenvalues, and \emph{eo ipso} infinitely many unstable modes around
$u_S(r)$. We found numerically that the "most unstable" mode has the
eigenvalue $k^2=-0.061306$. The spectrum of perturbations around the rescaled
solutions $u_S^a(r)$ is obtained by scaling $v(r) \rightarrow v(kr)$, $k^2 \rightarrow
a^2 k^2$.

To summarize, there exists the static solution $u_S(r)$ (and the continuous family
of its rescalings $u_S^a(r)$) which has infinite energy and infinitely many unstable modes.
Can such a beast play any role in dynamical evolution? The answer is not clear to us.
The point is that both the infinite energy and infinite instability of $u_S(r)$ have an origin in
the far-field behavior so it does not seem impossible that solutions $u_S^a(r)$ truncated at some
radius appear as local attractors\footnote{In a recent paper~\cite{lhi} Liebling, Hirschmann,
and Isenberg claim to have seen the solutions $u_S^a(r)$ at the threshold for singularity formation in the evolution
of very special initial data of noncompact support. We have serious misgivings about this result,
in particular we do not understand the discussion of "critical" solutions which are not intermediate
attractors.}.

 An alternative way of looking at this issue is to consider
solutions of
(\ref{static}) in a finite region $r\leq R$, that is harmonic maps from a
ball $B^3(R)$ into $S^3$. In the Dirichlet case, $u(R)=c$, the
number of such  solutions
 depends on the
value of a constant $c$ -- this was discussed in detail by J\"ager
and Kaul~\cite{jk}. In the Neumann case, $u'(R)=0$, which might be
more relevant in dynamics, there exists a countable family of finite
energy regular solutions $u_k(r)$. They are given by
\begin{equation}\label{nem}
  u_k(r)=u_S^{a_k}(r) \quad \text{with} \quad a_k=\frac{r_k}{R},
\end{equation}
where $r_k$ is the $k$-th Neumann point of $u_S(r)$, that is a point
where $u_S'(r_k)=0$ ($k=1,2,\ldots$). By construction
 the solution $u_k(r)$ has $k-1$ extrema on $r\in (0,R)$.
 By the same Sturm-Liouville theory argument
as above, one can show that a truncated solution $u_k$ has exactly
$k-1$ instabilities so a priori it might appear as a codimension $(k-1)$
local attractor in the dynamical evolution.

In passing we remark that a similar structure of static solutions arising in the chemotaxis problem
was discussed by Brenner et al.~\cite{brenner}.

\section{Numerical results}
In this section we describe the results of our numerical simulations
of the Cauchy problem (\ref{eq})-(\ref{data}). The main goal of these simulations
was to identify the generic final states of evolution (stable
attractors) and determine the boundaries between their basins of
attraction. We emphasize that
  the convergence to attractors (which is due to radiation of energy
to infinity) is always meant in a local sense. Before going into
details, we would like to say a few words about the numerical
techniques we employed. The simulations were performed by two
different finite difference methods. The first method
was based on an adaptive  mesh refinement algorithm. This code allowed us
to probe the structure of solutions near the singularity
with good resolution. The second method, designed specially to
study the convergence to self-similar solutions, solved the Cauchy
problem for Eq.(\ref{rho-tau}) on a fixed grid. In this
case there was no need of mesh refinement because the convergence to self-similar
profiles is a smooth process in similarity variables.
The main difficulty of using similarity variables is that we do not know
the blowup time $T$ in advance, which means that we have to deal with
the gauge mode instability. To suppress this instability (that is, to guess a blowup
time $T$) we
fine-tuned an extra parameter in the initial data. The fact that the
two independent numerical techniques generated basically the same outputs
makes
us feel confident about our results.

We remind that initial data of finite energy can be classified according to the
topological degree of the wave map at a fixed time (which is a map
from topological $S^3$ into $S^3$). Since the degree is preserved by
evolution, the Cauchy problem breaks into infinitely many topological
sectors. The nonzero degree data are not small by definition, and we
conjecture that they always develop singularities. Thus, from the
point of view of studying the threshold for singularity formation,
only degree zero data are interesting so most of our discussion is
focused to such data. A typical example is an  ingoing
"gaussian"
\begin{equation}\label{gauss}
  u(r,0)=\phi(r)=A\: r^3
  \exp\left[-\left(\frac{r-r_0}{s}\right)^4\right],\quad
  u_t(r,0)=\psi(r)=\phi'(r).
\end{equation}
In agreement with rigorous results of \cite{kov, sid} we found that if the initial data
are sufficiently small then the solution disperses, that is it
converges uniformly on any compact interval to the "vacuum" solution
$u=0$. In contrast, large initial data develop singularities in
finite time -- this manifests itself in an unbounded growth of the
gradient of solution at $r=0$. The precise character of blowup will
be described below.
\subsubsection*{Threshold behavior}
In order to determine the boundary between  two generic asymptotic
states of evolution, dispersion and collapse, we considered
  the evolution of various interpolating one-parameter families of
initial data $(\phi(r,p),\psi(r,p))$, that is such families that the
corresponding solutions exist globally if the parameter $p$ is small
and  blow up if the parameter $p$ is large.  Along each interpolating
family
 there must
exist a critical parameter value $p^*$ which separates these two
regimes. Given two values $p_{small}$ and $p_{large}$, it is
straightforward (in principle) to find $p^*$  by bisection. Repeating
this for many different interpolating families of initial data one
obtains a set of critical data  which by construction belongs to the
threshold. In order to figure out the structure of the threshold one
needs to determine the flow  of critical data.
 The precisely critical data cannot be prepared numerically but in practice it is
  sufficient to follow the
  evolution of marginally critical data.  We  found that the flow of such data
   has a transient phase when it seems to approach
  the self-similar solution $f_1(r/(T-t))$ for some $T$ (see Fig.~4). This
  behavior is universal in the sense that it is independent of the
  family of initial data; only the parameter $T$ depends on the data.

  This kind of behavior can be naturally explained as follows.
  As we showed above, the self-similar solution $f_1$ has exactly one
  unstable mode (apart from the gauge mode) -- in other words the
  stable manifold of this solution, $W_S(f_1)$, has codimension one and therefore
  generic one parameter families of initial data do intersect it. The
  points of intersection correspond to critical initial data that
  converge asymptotically to $f_1$.
  The marginally
critical data, by continuity, initially remain close to $W_S(f_1)$
and approach $f_1$ for intermediate times but eventually are repelled
from its vicinity  along the one-dimensional unstable manifold (see Fig.~5).
Within this picture the universality of marginally critical dynamics
in the intermediate asymptotics follows immediately from the fact
that the same unstable mode dominates the evolution of all solutions. More precisely,
the evolution of marginally critical solutions in the
  intermediate asymptotics can be approximated as
  \begin{equation}\label{inter}
  u(t,r) = f_1(\rho) +
   c(p) e^{\lambda \tau} v(\rho)/\rho + \mbox{decaying
  modes}
,
  \end{equation}
  where $\rho=r/(T-t)$, $\tau=-\ln(T-t)$, and
  $\lambda=\lambda^{(1)}_1 \approx 6.3336$.
  The small constant $c(p)$, which is the only vestige
   of the initial data, quantifies an admixture of the unstable mode -- for
   precisely critical data $c(p^*)=0$. The "lifetime" $\tau^*$ of the transient
    phase during which
   the linear  approximation (\ref{inter}) is valid
  is determined by the time in which the unstable mode grows to a
  finite size, that is  $c(p) e^{\lambda \tau^*} \sim O(1)$. Using
  $c(p)\approx c'(p^*) (p-p^*)$, this gives  $\tau^*\sim
  -\frac{1}{\lambda} \ln|p-p^*|$.
\begin{figure} [!ht]
\centering
\includegraphics[width=\textwidth]{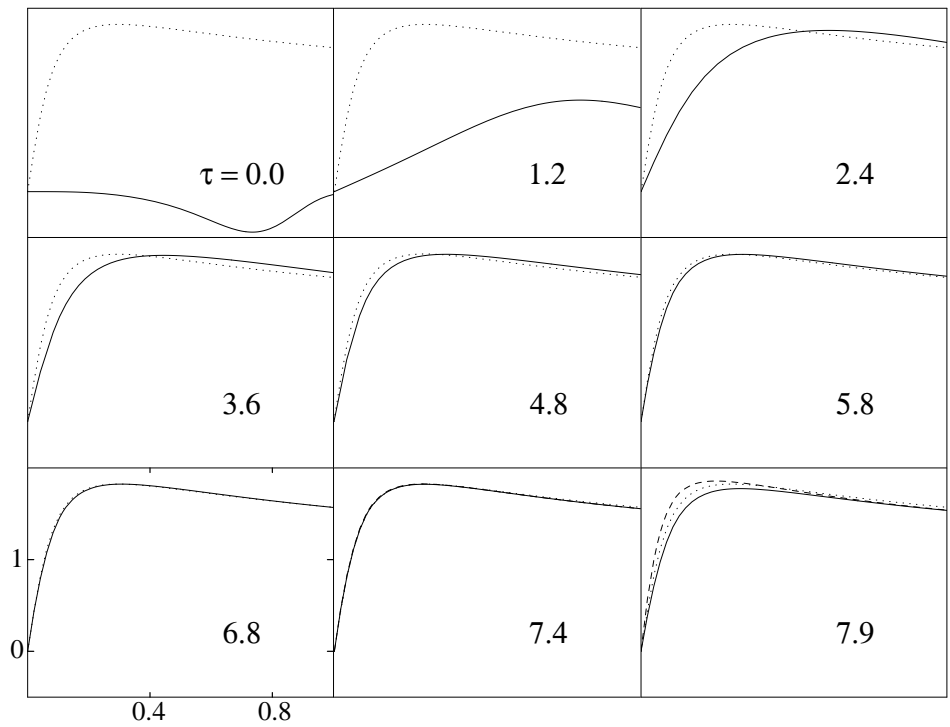}
\caption{The plot of $u(\tau,\rho)$ against $\rho$ from the evolution
(in similarity variables) of two marginally critical  gaussian-type
initial data of the form (\ref{gauss}), one subcritical (solid line) and one supercitical
(dashed line). These data are identical, except for the amplitudes
which differ  by $10^{-17}$,
 so the solutions practically coincide until the last frame. The
influence of the gauge mode instability is minimized by fine-tuning
the width of the gaussian. The convergence to the self-similar profile
$f_1(\rho)$ (dotted line) is clearly seen. In the last frame the two
solutions depart from the intermediate attractor in the
opposite directions. }
\end{figure}
\begin{figure}[!ht]
\centering
\includegraphics[width=0.9\textwidth]{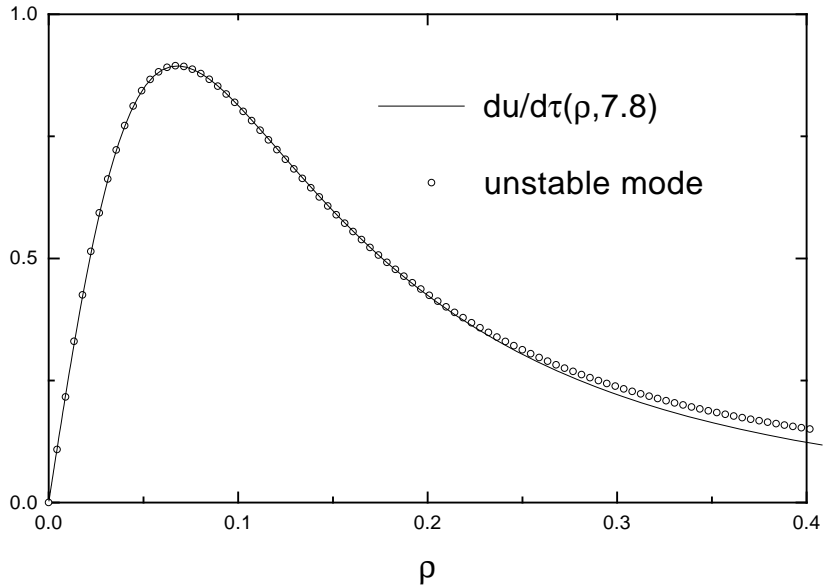}
\caption{Departure of the supercritical solution shown in Fig.~4 from
the intermediate attractor. The $\tau$-derivative of the solution is
shown to coincide (for small $\rho$) with the suitably normalized unstable mode around
$f_1$.}
\end{figure}
\subsubsection*{Universality of blowup}
We address now the question: what is the shape of solutions as they
approach the singularity?
We consider first the kink-type initial data of degree one, for example
$u(0,r)=\phi(r)=\pi \tanh(r/s)$. We found that such data always blow up in a finite time $T$
and the asymptotic shape of solution near $r=0$ approaches the TS solution $f_0(r/(T-t))$ as $t\rightarrow T^-$.
In this sense the singularity formation can be considered as local convergence to $f_0$.
This is shown in Figs.~6 and 7.
\begin{figure}[!ht]
\centering
\includegraphics[width=\textwidth]{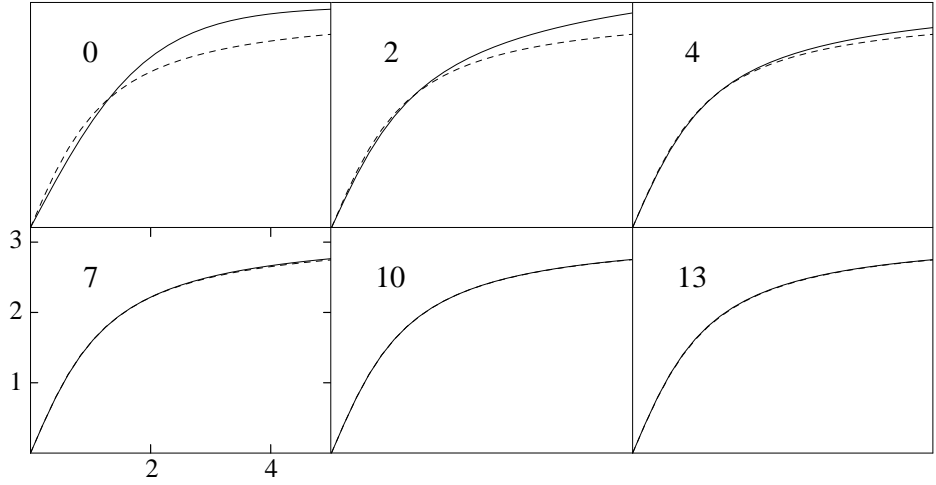}
\caption{The evolution of kink-type initial data $u(0,\rho)=\pi
\tanh(\rho/s)$ in similarity variables. The solution (solid line) converges to the
Turok-Spergel solution $f_0(\tfrac{\rho}{1+\epsilon e^{\tau}})$ (dotted line). By fine-tuning
the parameter $s$,
an admixture of the gauge mode instability quantified by $\epsilon$ was made very small, $\epsilon=-0.0085 \:e^{-13}$,
so for times $\tau<13$ the profile  $f_0(\tfrac{\rho}{1+\epsilon e^{\tau}})$ is practically static.}
\end{figure}
\begin{figure}[!ht]
\centering
\includegraphics[width=0.9\textwidth]{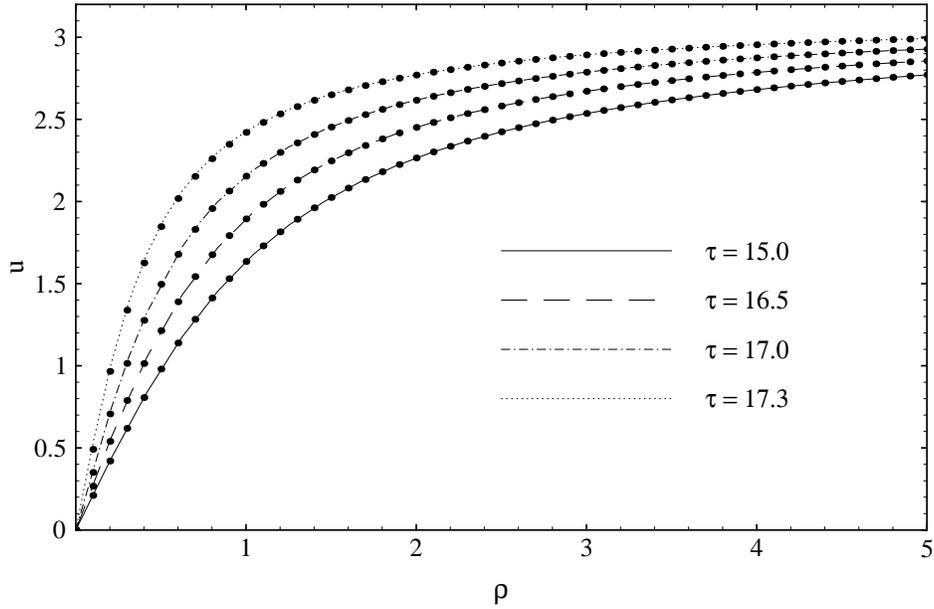}
\caption{The same solution as in Fig.~6 at later times when the gauge mode instability shows up. The solution follows the
 moving attractor $f_0(\tfrac{\rho}{1+\epsilon e^{\tau}})$ (dots).}
\end{figure}

We have observed the same behavior in other topological sectors,
in particular in the case of supercritical degree zero data. In Figs.~8 and 9 we show the formation of a self-similar
singularity in the collapse of slightly perturbed solution $f_1$.
\begin{figure} [!ht]
\centering
\includegraphics[width=0.8\textwidth]{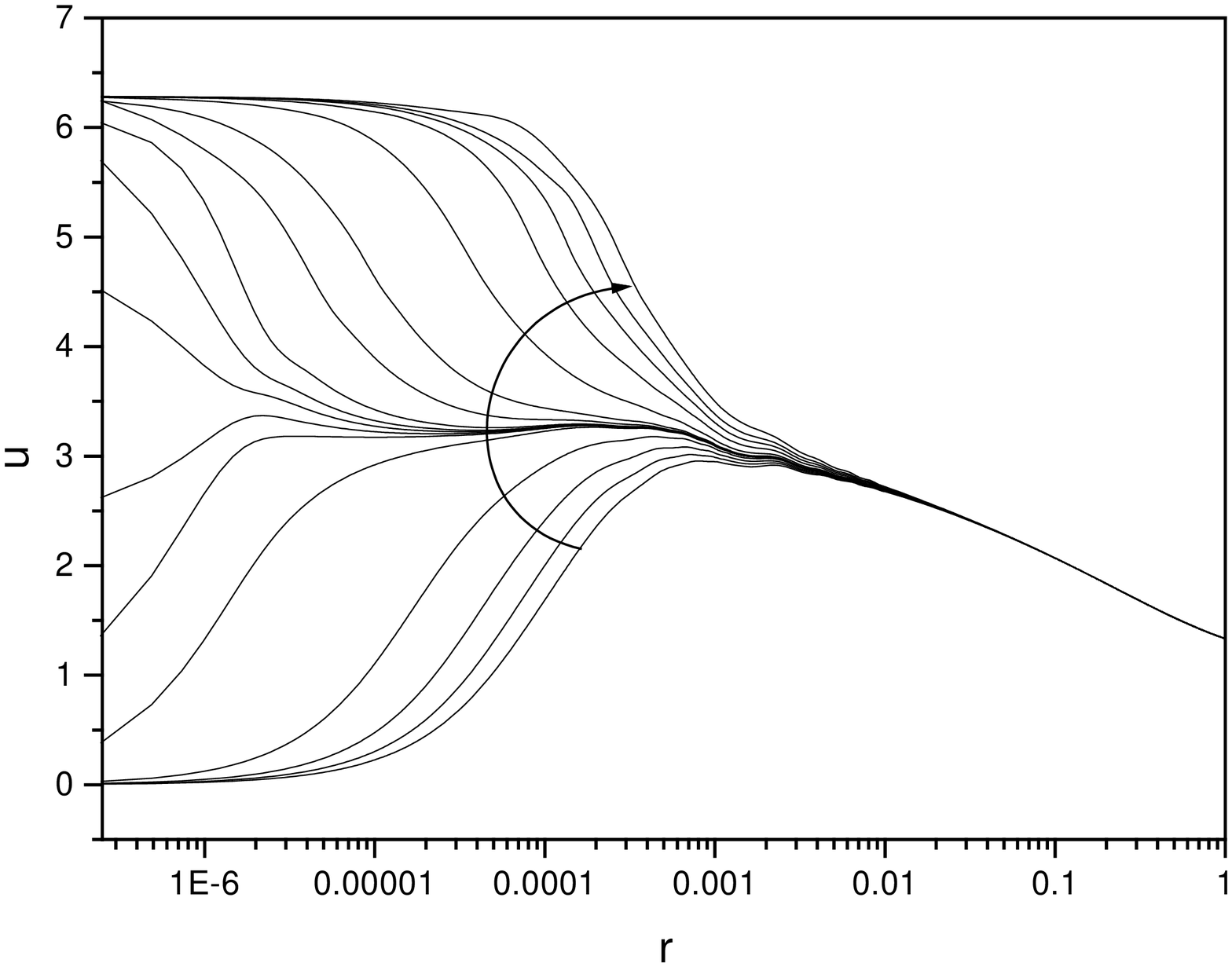}
\caption{The last stages ($|T-t|<10^{-5}$) of collapse of marginally supercritical
initial data (the solution $f_1$ was gently "pushed" towards collapse). The arrow indicates
the direction of increasing time. The
rapidly evolving inner region and the almost frozen outer region can be
clearly distinguished -- this is a typical situation in the formation of a localized singularity.
The numerical solution passes through the blowup in an
almost continuous manner -- only the point $u(r=0,t)$ jumps from $0$
to $2\pi$ as $t$ crosses $T$.}
\end{figure}
\begin{figure}[!ht]
\centering
\includegraphics[width=0.8\textwidth]{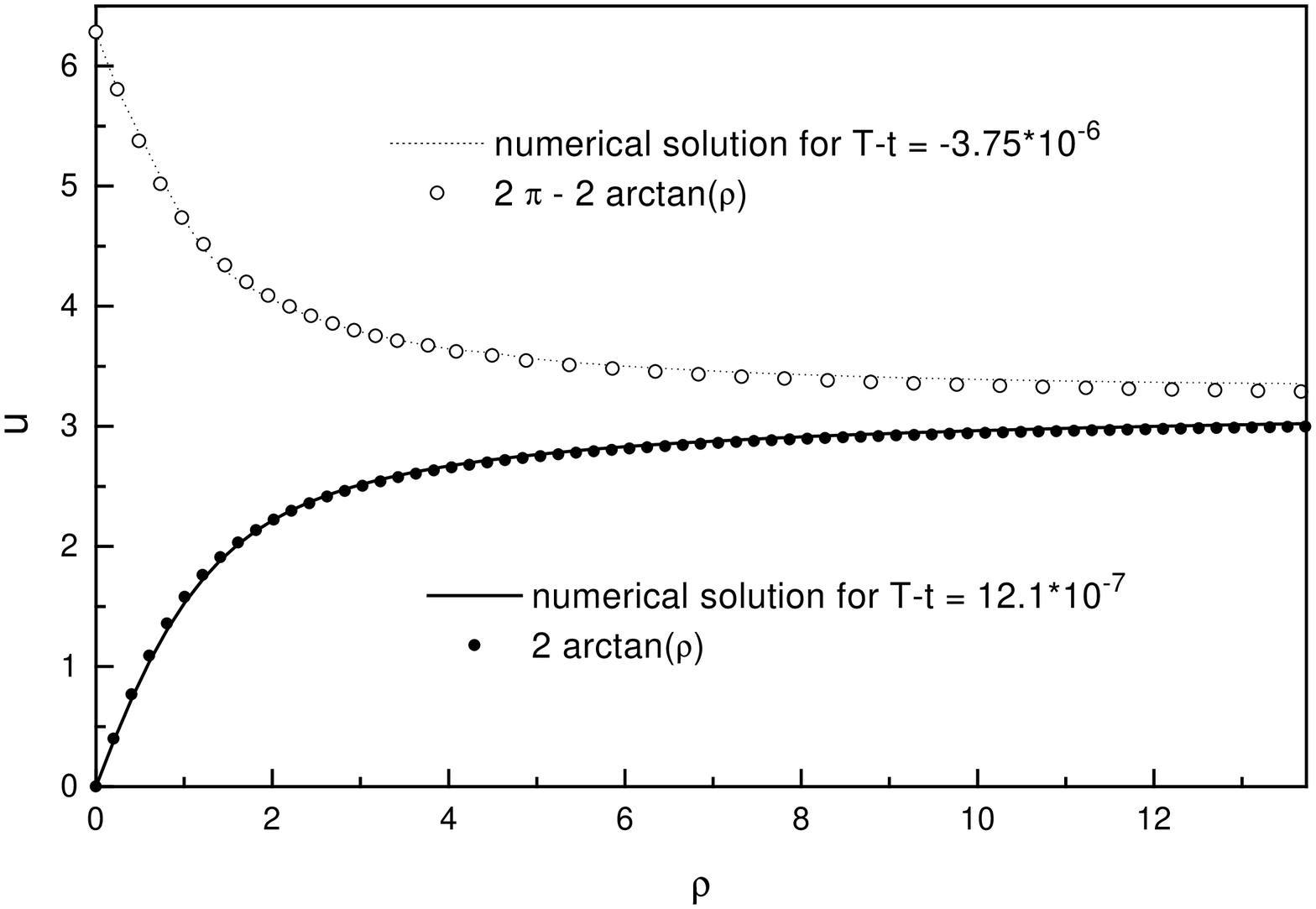}
\caption{Evidence of universal self-similarity of blowup. The
profiles just before and after the blowup are shown to coincide  with the
Turok-Spergel solution and its reflection.}
\end{figure}
On the basis of these numerical observations, we
conjecture that for a large set of solutions which blow up in finite time,
the asymptotic shape
near the singularity is given by the self-similar solution $f_0$. More precisely, for such solutions
there
exists a time $T$ such that
\begin{equation}\label{shape}
  u(r,t) \rightarrow f_0\left(\frac{r}{T-t}\right) \quad \text{as} \quad t
  \rightarrow T^-
\end{equation}
inside the past light cone of the point $(T,0)$.

Note that in the case of self-similar blowup the energy does not concentrate
at the singularity; in fact the energy  inside the past light cone of the point $(T,0)$ decreases
linearly with $T-t$. This suggests that the solutions can be continued beyond
the blowup time. Indeed, in Figs.~8 and 9 we show solutions just after the blowup.
At $r=0$ the solution $u(t,0)$  jumps from $0$ to $2\pi$ as $t$ crosses $T$. Since $u=0$ and $u=2\pi$
 correspond geometricaly to the same point, namely the north pole of $S^3$, the
solution passing through the singularity remains smooth everywhere, except at one point $(0,T)$. Moreover,
the solution retains the self-similar profile at least for some time after the blowup.
\section{Conclusions}
We have studied the Cauchy problem for corotational wave maps from  $3+1$ Minkowski spacetime into the 3-sphere.
We found that self-similar solutions play a special role in the dynamical evolution.
The stable self-similar solution (Turok-Spergel solution) determines the asymptotic profile
of solutions that blow up in finite time. The self-similar solution with one instability
plays the role of a critical solution, that is, its stable manifold separates solutions
that blow up from solutions that disperse. Of course, it is impossible to explore numerically
the whole phase space, so the complete picture of singularity formation and critical behavior  might
be richer than the one sketched here.
In particular our analysis leaves open the question about the role of a family of static solutions.
Although we have not systematically investigated the nontrivial topological sectors of the model, we anticipate
a rich phenomenology of singularity formation for solutions with high degree; for example we have observed such
solutions evolving (in a weak sense) through a sequence of blowup times $T_i$.

In our opinion the most interesting open  question is: why the large data solutions
become self-similar near the singularity? We think that this problem should be approached in
similarity variables in which the problem of blowup translates into a question of asymptotic
behavior as $\tau \rightarrow \infty$. Note that
the evolution equation expressed in similarity variables (\ref{rho-tau}) resembles the wave equation
 with damping.
It is thus natural to seek a Lyapunov functional, that is a functional that decreases in time on solutions.
If such a functional exists then its minima are the candidates for generic asymptotic states of evolution,
while its saddle points are the candidates for positive codimension attractors. Now, the self-similar solutions $f_n$
restricted to the interval $\rho\in[0,1]$
are the critical points of the functional
\begin{equation}\label{lyapunov}
K[u] = \frac{1}{2} \int_0^1 \left(\rho^2 u_{\rho}^2 - \frac{2 \cos^2(u)}{1-\rho^2} \right) d\rho.
\end{equation}
Although we were unable to show that this is a Lyapunov functional for Eq.(\ref{rho-tau}), we
believe that the mechanism suggested here is responsible for asymptotic self-similarity of
blowup.

\section*{Acknowledgement}
The results of this work were announced by one of us
in~\cite{needs} and~\cite{ja}. Later, there appeared a paper by Liebling, Hirschmann, and Isenberg on
the same subject~\cite{lhi}, in which  criticality of the self-similar solution $f_1$ was also
 observed.\\
This research was supported in part by the KBN grant 2 P03B 010 16.

\end{document}